\documentclass[aps,prd,10pt,twocolumn,preprintnumbers,superscriptaddress,bibnotes,nofootinbib,letterpaper,longbibliography]{revtex4-2}
\pdfoutput=1
\usepackage{graphicx} 
\usepackage{placeins} 
\usepackage[dvipsnames]{xcolor}
\usepackage{scalefnt}
\usepackage{booktabs}
\usepackage{xspace}   
\usepackage{hyperref}
\usepackage{amsmath}
\usepackage{array}
\usepackage[normalem]{ulem} 

\usepackage{enumitem} 

\newcommand{\myscalefont}[1]{{\scalefont{0.8}#1}}
\newcommand{\GENIE}{G\myscalefont{ENIE}\xspace}
\newcommand{\nuwro}{N\myscalefont{uWro}\xspace}

\newcolumntype{C}[1]{>{\centering\arraybackslash}m{#1}}

\def\Fermilab{Theory Division, Fermilab, P.O.\ Box 500, Batavia, IL 60510, USA}

\newcolumntype{Y}[1]{>{\centering\arraybackslash}m{#1}}

\begin{document}

\title{Improving Neutrino Oscillation Measurements through Event Classification}

\author{Sebastian A. R. Ellis}
\email{sebastian.ellis@kcl.ac.uk}
\affiliation{King’s College London, Strand, London, WC2R 2LS, United Kingdom}
\affiliation{Département de Physique Théorique, Université de Genève,
24 quai Ernest Ansermet, 1211 Genève 4, Switzerland}

\author{Daniel C. Hackett}
\email{dhackett@fnal.gov}
\affiliation{\Fermilab}

\author{Shirley~Weishi~Li}
\email{shirley.li@uci.edu}
\affiliation{Department of Physics and Astronomy, University of California, Irvine, CA 92697, USA}

\author{Pedro A. N. Machado}
\email{pmachado@fnal.gov}
\affiliation{\Fermilab}
 
\author{Karla Tame-Narvaez}
\email{karla@fnal.gov}
\affiliation{\Fermilab}

\date{November 14, 2025}

\begin{abstract}
Precise neutrino energy reconstruction is essential for next-generation long-baseline oscillation experiments, yet current methods remain limited by large uncertainties in neutrino–nucleus interaction modeling. 
Even so, it is well established that different interaction channels produce systematically varying amounts of missing energy and therefore yield different reconstruction performance--information that standard calorimetric approaches do not exploit. 
We introduce a strategy that incorporates this structure by classifying events according to their underlying interaction type prior to energy reconstruction. 
Using supervised machine-learning techniques trained on labeled generator events,
we leverage intrinsic kinematic differences among quasi-elastic scattering, meson-exchange current, resonance production, and deep-inelastic scattering processes. 
A cross-generator testing framework demonstrates that this classification approach is robust to microphysics mismodeling and, when applied to a simulated DUNE $\nu_\mu$ disappearance analysis, yields improved accuracy and sensitivity at the 10-20\% level. 
These results highlight a practical path toward reducing reconstruction-driven systematics in future oscillation measurements.
\end{abstract}

\preprint{FERMILAB-PUB-25-0618-T, UCI-HEP-TR-2025-11}

\maketitle


\section{Introduction}
\label{sec:intro}

Next-generation long-baseline neutrino oscillation experiments like DUNE and Hyper-Kamiokande aim to discover leptonic charge-parity (CP) violation, measure neutrino mass ordering, and probe physics beyond the Standard Model. 
Unprecedented experimental precision is required to achieve these goals.
Consequently, systematic uncertainties must be controlled to a much more stringent level than previously~\cite{DUNE:2020ypp,Hyper-Kamiokande:2018ofw}. 

A dominant source of systematic uncertainties is related to neutrino energy reconstruction. 
Since neutrinos are electrically neutral, they cannot be directly detected; instead, the incident neutrino energy must be inferred from the kinematic properties of charged particles produced in neutrino-nucleus interactions~\cite{Friedland:2018vry}. 

Experiments with multi-GeV beam energies, such as NOvA and DUNE, employ a calorimetric energy reconstruction strategy~\cite{DUNE:2020ypp,NOvA:2021nfi}. 
This method operates on the principle of energy conservation: the incoming neutrino energy is reconstructed by summing the measured energies of all detected final-state particles produced in the interaction. 
However, this initial sum systematically underestimates the true neutrino energy due to several factors: energy bound in the mother nucleus, undetected neutral particles such as neutrons, and charged particles with energies below detector thresholds. 
To account for these missing-energy contributions, correction factors derived from event generator simulations of neutrino-nucleus interactions are applied to the measured visible energy. 
The accuracy of these corrections depends critically on the precision of the underlying cross-section models used in event generators~\cite{Ankowski:2015kya,Friedland:2018vry,CLAS:2021neh,Coyle:2025xjk}.

Due to the lack of scale separation between GeV neutrinos and the QCD energy scale that dictates nuclear physics dynamics, the theoretical computation of such neutrino-nucleus cross-sections is challenging~\cite{NuSTEC:2017hzk,Kronfeld:2019nfb,Ruso:2022qes}.
These interactions transfer $Q^2 \sim 1-10~\text{GeV}^2$ to the nucleus, exciting a variety of hadronic degrees of freedom including nucleons, baryonic resonances, and partons~\cite{Bloom:1970xb,Rein:1980wg,Bodek:2002vp,Benhar:2005dj,Ankowski:2005wi,Drechsel:2007if,Bodek:2010km,Kamano:2013iva,Nakamura:2015rta,Kabirnezhad:2017jmf,Xie:2023suk}. 
Neutrino event generators attempt to model these dynamics by relying on a combination of models, each of which is designed to handle the appropriate hadronic degrees of freedom in the corresponding region of the collision phase space~\cite{Andreopoulos:2009rq,Hayato:2009zz,Buss:2011mx,Golan:2012rfa,Gallmeister:2016dnq,Isaacson:2022cwh}.
Quasi-elastic scattering (QE) treats the neutrino as interacting with entire nucleons.
Resonance production (RES) accounts for neutrino interactions exciting baryonic resonances.
Deep-inelastic scattering (DIS) treats the nuclei at the partonic level.
Finally, since neutrino-nucleus scattering occurs within a nuclear medium, there is the possibility that correlated nucleon pairs are excited through meson-exchange currents (MEC).
While not an exhaustive listing, these four mechanisms are presently considered as the dominant set relevant for accelerator neutrino experiments. 
Each of these models for neutrino-nucleus interactions can intuitively be associated to a reasonably well-defined energy transfer regime.
Nevertheless, there is typically significant overlap between them, leading to theoretical and ultimately experimental uncertainties.

Given these theoretical complexities, numerous efforts have focused on mitigating the impact of reconstruction errors through refined analysis techniques. 
These approaches can be broadly categorized into two strategies. 
The first strategy aims to improve the reconstruction algorithm itself. 
For instance, several groups have explored using convolutional neural networks to improve neutrino energy reconstruction, leveraging their ability to identify complex patterns in detector data that correlate with true energy~\cite{Yu:2019yuj, Liu:2020pzv, Shmakov:2023jms, IceCube:2025jmv, Baldi:2018qhe}. 
A second, more conceptually intuitive strategy involves classifying events to account for reconstruction differences. 
A pioneering example of this was developed by the NOvA collaboration in Refs.~\cite{NOvA:2018gge,Vinton:2018aqq}, whose analysis divided their event sample into four quartiles after reconstruction based on each event's estimated energy resolution quality, and thereby improving precision in oscillation analyses. 

In this paper, we introduce a new classification strategy to improve oscillation analyses: classifying events using their intrinsic kinematic properties rather than reconstruction-dependent metrics.
Our approach leverages the inherent kinematic differences between interaction types.
For example, QE events typically have less missing energy and can be reconstructed well, whereas DIS events have significant missing energy and worse resolution. 
Standard calorimetric methods treat all events homogeneously and are thus prone to conflating systematic uncertainties associated with energy reconstruction.
To overcome this, we classify events according to their underlying interaction type prior to energy reconstruction, and then employ an analysis strategy which treats these classes separately.

To classify neutrino events, we use supervised machine learning (ML) techniques, which require correct interaction-type labels for training.
This information is unavailable for real-world experimental data; event classifiers must thus be trained on event-generator data where the interaction type for each event is known. 
This raises the concern that the classifier will learn generator-specific traits as opposed to generic features of the kinematics resulting from the microphysics of the different interactions. 
We assess this directly using mock data from multiple generators, with one generator simulating the role of experimental data in the analysis---except with labels that can be used to quantify how well the classifier generalizes.
In this testing framework, we are able to demonstrate an approach that is robust against microphysics mismodeling and is thus suitable for application to experimental data.

To estimate the impact of our method on a neutrino oscillation measurement, we simulate a toy DUNE $\nu_\mu\to\nu_\mu$ disappearance analysis. 
We compare analyses performed with and without event classification to quantify the improvement, and we use either the same or different generators for the fit and mock experimental data to study the effects of microphysics mismodeling.
We find encouraging improvements in the sensitivities and accuracies of atmospheric oscillation parameters.
Furthermore, the improvements are robust against interaction mismodeling, demonstrating the method's potential for real-world experimental applications.

The remainder of this paper is organized as follows. In Section~\ref{sec:interaction}, we discuss the physics of neutrino-nucleus interactions and how different event types lead to different neutrino energy resolution. 
In Section~\ref{sec:classification}, we describe the classification framework for identifying neutrino interaction types based on final-state observables, and examine its dependence on generator modeling. 
In Section~\ref{sec:oscillation}, we integrate the classified events into a simplified oscillation analysis to evaluate the improvement on oscillation measurements.
We conclude in Section~\ref{sec:conclusions} with a discussion of the broader implications for event selection, cross-section tuning, and systematic uncertainty reduction in future experiments.


\section{Neutrino-nucleus interactions and their interplay with neutrino energy measurement}
\label{sec:interaction}

This section describes the underlying physics that causes different neutrino interaction types to exhibit distinct energy resolution characteristics. 
As our focus is on DUNE, we center the discussion on the liquid argon time projection chamber (LArTPC) technology. We also detail how we simulate neutrino interactions in a LArTPC. 


\begin{figure}[t]
\begin{center}
\includegraphics[width=\linewidth]{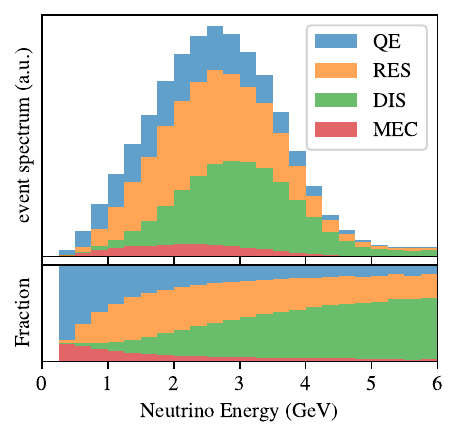}
\caption{Neutrino event spectrum in the DUNE near detector split according to interaction type. Quasi-elastic (QE), resonant (RES), deep inelastic scattering (DIS), and meson exchange currents (MEC) are shown from top to bottom in each panel.
We focus on neutrino energies relevant to oscillation physics.}
\label{fig:spectrum}
\end{center}
\end{figure}

\subsection{Neutrino-nucleus interactions}

Throughout this work, we simulate neutrino interactions using two open-source event generators: \GENIE version 3.04.00 with the \texttt{G18\_02a\_00\_000} tune~\cite{Andreopoulos:2009rq,Andreopoulos:2015wxa} and \nuwro version 21.09.2~\cite{Golan:2012rfa}. 
Since our mock analysis is for DUNE, we use the DUNE $\nu_\mu$ flux spectrum~\cite{DUNE_fluxes} for event generation.
Below, we outline the modeling of neutrino-nucleus interactions and the definition of interaction types in these generators, and note important caveats.

In the neutrino beam energy range of 0.5--5~GeV, neutrino-nucleus cross sections can typically be modeled using the ``impulse approximation,'' where the nucleus is treated as a collection of independent nucleons~\cite{Benhar:2005dj,Nieves:2011pp}. 
The interaction can be understood as follows: first, in the \textit{primary interaction}, a neutrino interacts with a nucleon in the nucleus.
Then, the resulting outgoing particles propagate and undergo \textit{final-state interactions} until they exit the nucleus.

In the impulse approximation as implemented in \GENIE and \nuwro, the primary interaction comprises several distinct channels, illustrated in Fig.~\ref{fig:spectrum}. 
In quasi-elastic interactions (labeled as QE), a neutrino scatters off a single nucleon, producing a lepton and a nucleon: $\nu_\ell \, n \rightarrow \ell \, p$ for neutrinos and $\bar\nu_\ell \, p \rightarrow \ell \, n$ for antineutrinos~\cite{LlewellynSmith:1971uhs}. 
Neutrino scattering off a nucleon can also produce baryonic resonances (RES).
In these events, an on-shell baryon is produced that subsequently decays, typically to a nucleon and a meson, e.g., $\nu_\ell \, n \rightarrow \ell \, \Delta^+$ followed by $\Delta^+ \rightarrow n \,  \pi^+$~\cite{Rein:1980wg}. 
Given sufficiently high energy transfer, the neutrino resolves substructure in the nucleon, and interacts directly with partons (DIS)~\cite{Bodek:2002vp}.
These events typically yield multiple pions.
Finally, the neutrino can sometimes interact with a tightly bound nucleon pair rather than a single nucleon (MEC), e.g., $\nu_\ell \, n \, n \rightarrow \ell \, p \, n$~\cite{Martini:2009uj}.

The above are primary interaction types, each of which yields a predictable final state.
In principle, this would allow straightforward categorization of the interaction types.
However, the particles produced from primary interactions subsequently traverse and exit the nucleus, undergoing final-state interactions in the process.
These secondary interactions can lead to rescattering, further energy loss, and particle absorption or production.
This results in a non-trivial mapping from final state back onto the primary interaction type, complicating the categorization.
  
Note that event generators include an additional interaction type: coherent pion production, in which a neutrino elastically scatters off the entire nucleus, producing a pion but leaving the nucleus in its ground state, $\nu_\ell \,A\to\ell \,\pi^+A$~\cite{Rein:1982pf}. 
We exclude this channel in our analysis because it constitutes a subdominant contribution---accounting for 0.3\% of events in \GENIE and 0.9\% in \nuwro.

In addition, note that interaction types are treated as distinct in event generators, with their cross sections summed to obtain the total. 
In nature, however, such a clean separation of phase space may not exist. 
For example, high-mass resonance production and DIS appear intrinsically linked~\cite{Shifman:2000jv,Bigi:2001ys,Melnitchouk:2005zr,Lalakulich:2006yn,Lalakulich:2008tu}. 
Nevertheless, as we demonstrate below, event generator labels can be employed as a practical tool to differentiate events with different characteristic dynamics.


\subsection{Neutrino energy measurement}
 
A complete reconstruction analysis would require a full detector simulation, which is beyond the scope of this work. Instead, we adopt a simplified approach to model the detector response: after generating the final-state particles from the neutrino-nucleus interaction, we approximate their propagation and detection in a LArTPC by applying detection thresholds, energy smearing, and angular smearing.

First, we take the magnitude of the momentum of each particle and smear it randomly according to a Gaussian with fractional width $\sigma_p$. 
We recalculate the energy of the particle using the dispersion relation. 
If the kinetic energy of a particle is below the threshold $K_{\rm th}$, we assume the particle is not reconstructed and remove it from the event.
Finally, we smear the direction of each visible particle: we perform a polar rotation with respect to its momentum axis following a random Gaussian distribution with width given by the angular resolution $\sigma_\theta$, and then we randomize the azimuthal angle (still with respect to the original momentum direction) using a flat $[0,2\pi]$ prior.
Table~\ref{tab:detector-effects} summarizes the threshold and resolution values used in this work. Most values are taken from Ref.~\cite{DUNE:2015lol}.
\begin{table}[b]
    \centering
    \begin{tabular}{Y{1.5cm}|Y{1.5cm}|Y{1cm}|Y{0.5cm}} \hline
        Particle & $K_{\rm th}$ & $\sigma_p$ & $\sigma_\theta$ \\
        \hline
        $e,\mu,\gamma$   & 30 MeV & 2\% & $2^\circ$ \\ \hline
        $p,\pi^\pm,K^\pm$   & 100 MeV & 30\% & $5^\circ$ \\ \hline
    \end{tabular}
    \caption{Kinetic energy threshold and angular resolution values adopted in this work.
    }
    \label{tab:detector-effects}
\end{table}

In the following, we outline the physical motivations for these thresholds and smearing parameters and justify the specific input values adopted in our model. Our discussion is guided by the concept of missing energy---the portion of the initial neutrino energy that is not detectable in the final-state particles---which is crucial for understanding the quality of calorimetric energy reconstruction~\cite{Friedland:2018vry,Friedland:2020cdp}. We focus on the sources of this missing energy in a LArTPC, emphasizing fundamental theoretical limitations over practical detector effects like particle contamination.

Following a neutrino-nucleus interaction, most of the neutrino energy is typically carried by the outgoing lepton---a muon in the case of $\nu_\mu$ interactions. 
These muons produce long, straight tracks in the detector and can be reconstructed relatively well~\cite{MicroBooNE:2016pwy,DUNE:2020ypp}.
The remaining energy is distributed among the outgoing hadrons and contributes to nuclear binding energy release---a source of missing energy.

Among the outgoing hadrons, neutrons present a unique reconstruction challenge. 
Neutron-argon interactions can produce visible secondary particles such as protons or photons.
However, these secondaries often appear as isolated energy deposits rather than clear tracks, making them difficult to reliably associate with the primary interaction vertex.
Despite these complications, neutron reconstruction is an active research area~\cite{ArgoNeuT:2018tvi,CAPTAIN:2019fxo,Castiglioni:2020tsu,ARTIE:2022wqs,CAPTAIN:2022nzf,MicroBooNE:2024hun}. 
In this analysis, we adopt a conservative approach and treat all neutron energy as missing.

Additional neutral particles like photons and $\pi^0$ can also be produced.
Unlike neutrons, these are readily detectable as they produce electromagnetic showers that can be reconstructed with high precision. 
Charged hadrons have the poorest energy resolution among detectable particles.
This is primarily because they undergo reinteractions in the detector medium, producing further hadrons---including neutrons---and additional sub-threshold particles. We therefore assume a 30\% energy resolution and a detection threshold of 100~MeV for protons and charged pions~\cite{DUNE:2015lol,Friedland:2020cdp,MicroBooNE:2020akw}.

\begin{figure}[t]
\begin{center}
\includegraphics[width=\linewidth]{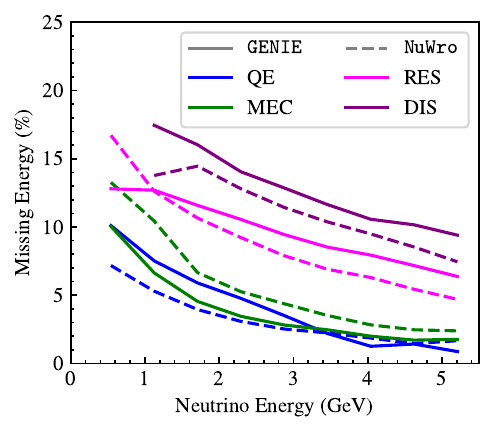}
\caption{Fraction of missing energy for  DUNE near detector events as a function of true neutrino energy. Different colors refer to the type of interaction, whereas \GENIE and \nuwro results are distinguished via solid and dashed lines. Note that the DIS cross section is negligible below 1 GeV.
}
\label{fig:Emiss}
\end{center}
\end{figure}

These considerations imply that the accuracy of the reconstructed energy will vary substantially across interaction types.
Figure~\ref{fig:Emiss} shows the fraction of missing energy, defined as $1-E_{\rm cal}/E_\nu$, as a function of neutrino energy $E_\nu$\footnote{The neutrino energy spectra in Figures~\ref{fig:spectrum} and \ref{fig:Emiss} are shown up to 6 GeV, as the DUNE flux is negligibly small beyond this energy. The oscillation analysis nonetheless extends to 10 GeV in calorimetric energy to ensure completeness.}.
The calorimetric energy is defined as 
\begin{equation}\label{eq:calo}
    E_{\rm cal} = E_\ell + E_{\rm mesons} +E_{\gamma} + K_{\rm p},
\end{equation}
where the first three terms is the total energy of charged leptons, detector-stable mesons, and photons, while the last term is the kinetic energy of protons.
The four dominant interaction types are shown separately, for both generators we study.
\GENIE typically estimates a higher missing energy than \nuwro.
However, both consistently predict a significant difference between the events with the most missing energy (DIS) and those with the least (QE).
Note also the clear separation in missing energy between soft processes (QE and MEC) and hard processes (RES and DIS).


\section{Classification}
\label{sec:classification}

In this section, we describe the strategy and implementation of the machine learning–based event classification framework. 
The primary goal is to assess whether supervised learning algorithms,
when applied to simulated neutrino events produced by Monte Carlo generators, 
can identify physically meaningful features corresponding to distinct interaction mechanisms. 
As an initial testbed, we consider a simple binary classification task---distinguishing QE, our hypothetical signal, from MEC, our putative background.
Subsequently, we extend to the four-way classification of separating QE, MEC, RES, and DIS events in a dataset.

Generalizability between different datasets is necessary to apply the method to real experimental data, for which labels from an exact model will not be available.
To evaluate whether classifiers are inadvertently learning generator-specific artifacts as opposed to meaningful physics, we train using data from one generator and test them on events from the other. 
This has the advantage that the simulated data we test on carries truth labels that allow us to assess how well the classifier has generalized.
Good generalization would be an indicator that when the classifier is used on real data without labels, we can be confident that the classification is driven by the underlying physics.

\begin{figure}[t!]
\begin{center}
\includegraphics[width=1\linewidth]{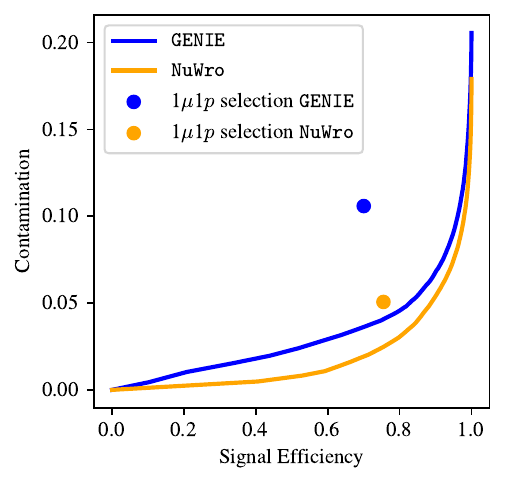}
\caption{Efficiency and contamination for the classification of QE and MEC events, assuming QE events as the signal. Results are shown for \GENIE (blue) and \nuwro (orange) events. The dots indicate benchmark points obtained using the cuts defined in the main text.}
\label{fig:1vs2}
\end{center}
\end{figure}

\subsection{Problem setup \& ML details}

For each of \GENIE and \nuwro, we simulate approximately $10^7$ neutrino events.
Labels (QE, MEC, RES, DIS) are assigned to each event according to the underlying interaction type as specified by the generator.
This ensures a dataset that is sufficiently large for the training of the interaction-type classifier, and is representative of the full range of possible final-state particle multiplicities.
Each dataset is split into training $(70\%)$ and testing $(30\%)$ subsets, using stratified sampling to preserve class balance.

The varying number of final-state particles in each event poses a challenge for architecture design:
each event corresponds to an input vector of a different dimension encoding the kinematics of the variable number of final-state particles.
To overcome this obstacle, we take the approach of computing statistical summaries of each event, and use these as fixed-size input vectors.
For each event we compute the mean and standard deviation of mass, energy, and momenta in each spatial direction, over all particles in the event.\footnote{For one-particle events, we define the standard deviation as zero.}
In addition, we include as input the number of particles per event, as well as the mean and standard deviation of each particle's contribution to the calorimetric energy of the event, i.e.~the kinetic energy of any protons and total energy of any leptons and mesons.

\begin{figure}[t!]
\begin{center}
\includegraphics[width=1\linewidth]{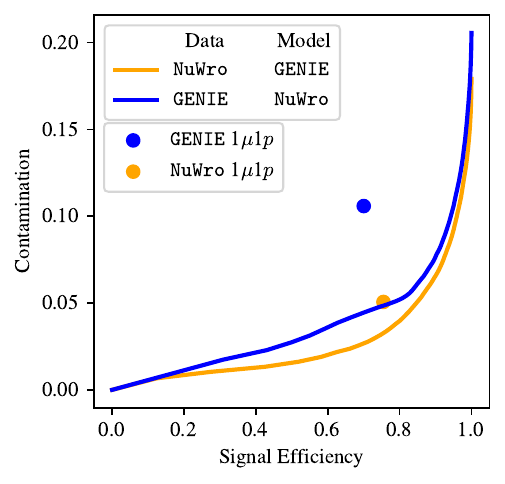}
\caption{Efficiency and contamination for the classification of QE and MEC events in the generalization scheme, i.e., with the data and model corresponding to different generators. As before, QE events are treated as the signal. Results are shown for \GENIE (blue) and \nuwro (orange) samples. The dots represent benchmark points obtained using cuts defined in the main text.}
\label{fig:cross_1vs2}
\end{center}
\end{figure}

The dimension of the output vector of a classifier model corresponds to the number of classes in the problem.
In this work, we consider binary classifier architectures which discriminate between two labels and thus have two output channels.
The output values are conventionally normalized to sum to one, thus admitting interpretation as class probabilities.
In the case of QE/MEC classification described in the next subsection, the binary classification is between two distinct interaction types.
In Section~\ref{sec:multi-class}, we extend these binary classifiers to multiple interaction types by employing a ``one-vs-rest'' strategy, described in detail below.

With the shape of the inputs and outputs defined, all that remains is to choose the model architecture and training scheme.
For all tasks, we use a multi-layer perception, i.e.~a fully connected Neural Network (NN)~\cite{Hornik:1989yye,Hinton:2012zps} implemented using the \texttt{Scikit-Learn} library~\cite{scikit-learn}. We use the default setup for \texttt{MLPClassifier} with the only changes being the NN architecture, consisting of two fully connected hidden layers with 64 and 32 neurons, respectively, and a training of 500 iterations.
Training is performed by optimizing the binary cross-entropy loss using the Adam optimizer~\cite{Kingma:2014vow}. 
Separate models are constructed and trained for each classification task discussed below. 

\subsection{QE vs.\ MEC}
\label{sec:binary-class}

For an initial exploration of this approach, we first build a classifier to distinguish between QE and MEC events, motivated by the current T2K oscillation analysis and MicroBooNE low-energy excess searches~\cite{T2K:2023mcm, MicroBooNE:2021pvo}.
Due to the low beam energy, T2K event samples are dominated by QE events, which provide clear kinematic reconstruction information. 
On the other hand, MEC events act as a contaminating background that degrades the energy reconstruction.
In addition to being motivated by experiment, this simple exercise serves to establish notation and conventions for the more complex multi-class case discussed below.
For this task, we restrict data to events with only QE or MEC interaction types.

For all binary classification tasks in this manuscript, we designate one class as the signal and the other as the background. 
A binary classifier produces a two-dimensional output $[1-p_\mathrm{signal}, p_\mathrm{signal}]$, which can be interpreted as class probabilities.
After training, 
a fixed classifier model still
defines a continuous space of classifiers via the choice of the threshold hyperparameter $\theta \in [0,1]$.
The threshold then defines how to assign the class label: outputs with $p_\mathrm{signal} > \theta$ are classified as signal, and those with $p_\mathrm{signal} \le \theta$ as background.
Different choices of $\theta \in [0,1]$ thus define a space of classifiers interpolating between two limiting cases: for $\theta=0$, all events are labeled as signal; for $\theta=1$, all are background.

To quantify the performance of a learned classifier, 
we compute the \textit{signal efficiency} and \textit{background contamination} 
as a function of the threshold hyperparameter $\theta$.
The efficiency and contamination are defined as
\begin{align}
\text{Efficiency} &= \frac{\text{TP}}{\text{TP} + \text{FN}},
\label{eq:eff}
\\[6pt]
\text{Contamination} &= \frac{\text{FP}}{\text{TP} + \text{FP}},
\label{eq:cont}
\end{align}
where
\begin{itemize}
    \item $\text{TP}$ (True Positives) are signal events correctly classified as signal,
    \item $\text{FN}$ (False Negatives) are signal events incorrectly classified as background, and
    \item $\text{FP}$ (False Positives) are background events incorrectly classified as signal.
\end{itemize}
An optimal classifier should simultaneously achieve both high signal efficiency and low background contamination so as to capture a majority of the true signal events while minimizing the number of background events in the signal sample.
Efficiency and contamination as a function of $\theta$ is a similar but more experimentally-relevant metric than the standard ``received operating characteristic (ROC)'' curve often used to characterize classifier performance.

For comparison with learned classifiers, we define a simple benchmark algorithm, ``$1\mu1p$'', based on kinematic cuts employed in common practice~\cite{T2K:2023mcm, MicroBooNE:2021pvo}.
This algorithm is: if the final state particles are \textit{exclusively} a muon and a proton, then the event is labeled QE; otherwise, it is labeled as background (in this case, MEC).
This zero-hyperparameter algorithm gives a single value of the efficiency and contamination for any given dataset.
We note, however, that this benchmark is intended only as a reference for a simple classical discrimination and does not represent a definitive standard.

Figure~\ref{fig:1vs2} shows the performance of the learned QE vs.\ MEC classifiers on each generator’s dataset. 
In this case, classifiers trained and tested on \nuwro generally achieve slightly lower contamination than those using \GENIE data. 
In both cases, some range of thresholds $\theta$ exists for which the learned classifier outperforms the benchmark analysis in \textit{both} efficiency and contamination (i.e., where the curve is both below and to the right of the benchmark point in Fig.~\ref{fig:1vs2}). 
However, perfect classification, i.e.\ 100\% signal efficiency with 0\% contamination, is not achieved for any $\theta$ with this simple setup.
Further improvements of the classifier model are conceivable.

We verify robustness against mismodeling by using the model trained on data from one generator to classify data from the other.
Our results are shown in Fig.~\ref{fig:cross_1vs2}.
As expected, imperfect training data (i.e.~the mismatch between the generator used for training and testing) leads to a degradation in classification performance. 
Notably, however, this degradation is small.
Comparing the results of Fig.~\ref{fig:cross_1vs2} with those of Fig.~\ref{fig:1vs2}, we see that the differences are minimal. 
This demonstrates that the classifier model is not simply learning generator-specific event characteristics, but is likely learning generic kinematic features associated to each event type. 
Given that, as described above, the input data consists of statistical summaries of kinematic features of events, it is perhaps not surprising that the classifier is not relying on spurious generator-dependent features to determine classification.
Nevertheless, the comparison between Figs.~\ref{fig:1vs2} and~\ref{fig:cross_1vs2} is an important validation of our approach.

\begin{figure*}[t!]
\begin{center}
\includegraphics[width=0.47\linewidth]{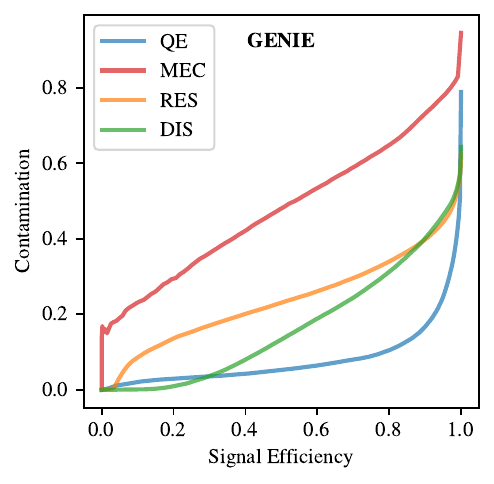}
\hfill
\includegraphics[width=0.47\linewidth]{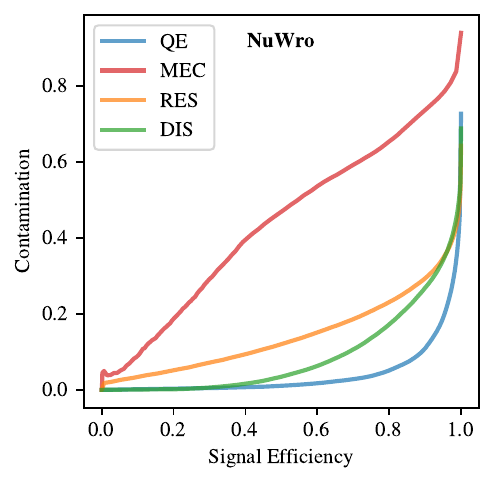}
\caption{Efficiency and contamination for \GENIE (left) and \nuwro (right) in a multi-class classification using the one-vs-rest strategy.}
\label{fig:bin_multi}
\end{center}
\end{figure*}


\subsection{Multi-class classification}
\label{sec:multi-class}

\begin{figure}[t!]
\begin{center}
\includegraphics[width=0.9\linewidth]{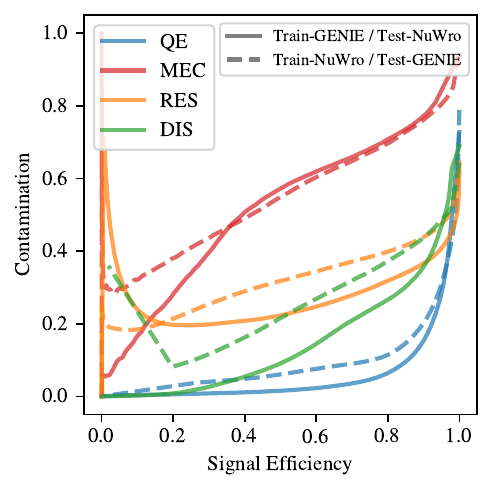}
\caption{Efficiency and Contamination in a multi-class classification using the one-vs-rest strategy. The solid/dashed lines represent the generalization scheme where data and model correspond to different generators.}
\label{fig:cross_multi}
\end{center}
\end{figure}

We now consider the more general multi-class case of labeling each event as QE, RES, MEC, or DIS.
To do so, we build a multi-class classifier using a ``one-vs-rest'' (OvR) strategy.
This classification strategy consists in iteratively designating one event type---e.g., QE---as signal, and performing a binary classification of signal and background until all four classes are identified.
The result is four sub-models, one that identifies QE vs.\ rest, another that identifies RES vs.\ rest, etc.
To predict the interaction type of an event, all four binary classifiers are queried; the truth label is taken as the one which classifies the event as signal with the highest confidence.
Such an OvR scheme allows defining efficiencies and contaminations as in the binary QE/MEC classification task above.
The four-class efficiencies and contaminations enable us to quantify how well a given event type (e.g., QE) can be discriminated against all others.

Figure~\ref{fig:bin_multi} compares classification performance on \GENIE and \nuwro events for all four OvR tasks.
The classifier performance is comparable for the two generators. 
However, the classifier performs somewhat better, especially for DIS and RES event types, when training and testing on \nuwro data than on \GENIE data.
For both generators, classifier performance is excellent at isolating QE, intermediate for RES and DIS, and comparatively poor for MEC.
The cause of the poor performance for MEC is unclear, but could be due to the relatively few events available to train the classifier and similar kinematics with QE. 
We note a spike in contamination at near-zero efficiency. As the decision threshold $\theta$ approaches 1, corresponding to labeling almost all events as background, only a small number of events are labeled as signal. As a result, large statistical fluctuations dominate both the numerator and denominator of Eqs.~\ref{eq:eff} and \ref{eq:cont}. 
This behavior is a finite-statistics artifact of the extreme threshold regime, and it does not affect the results of the oscillation analysis in Section~\ref{sec:oscillation}.

\begin{figure*}[t!]
\begin{center}
\includegraphics[width=1\linewidth]{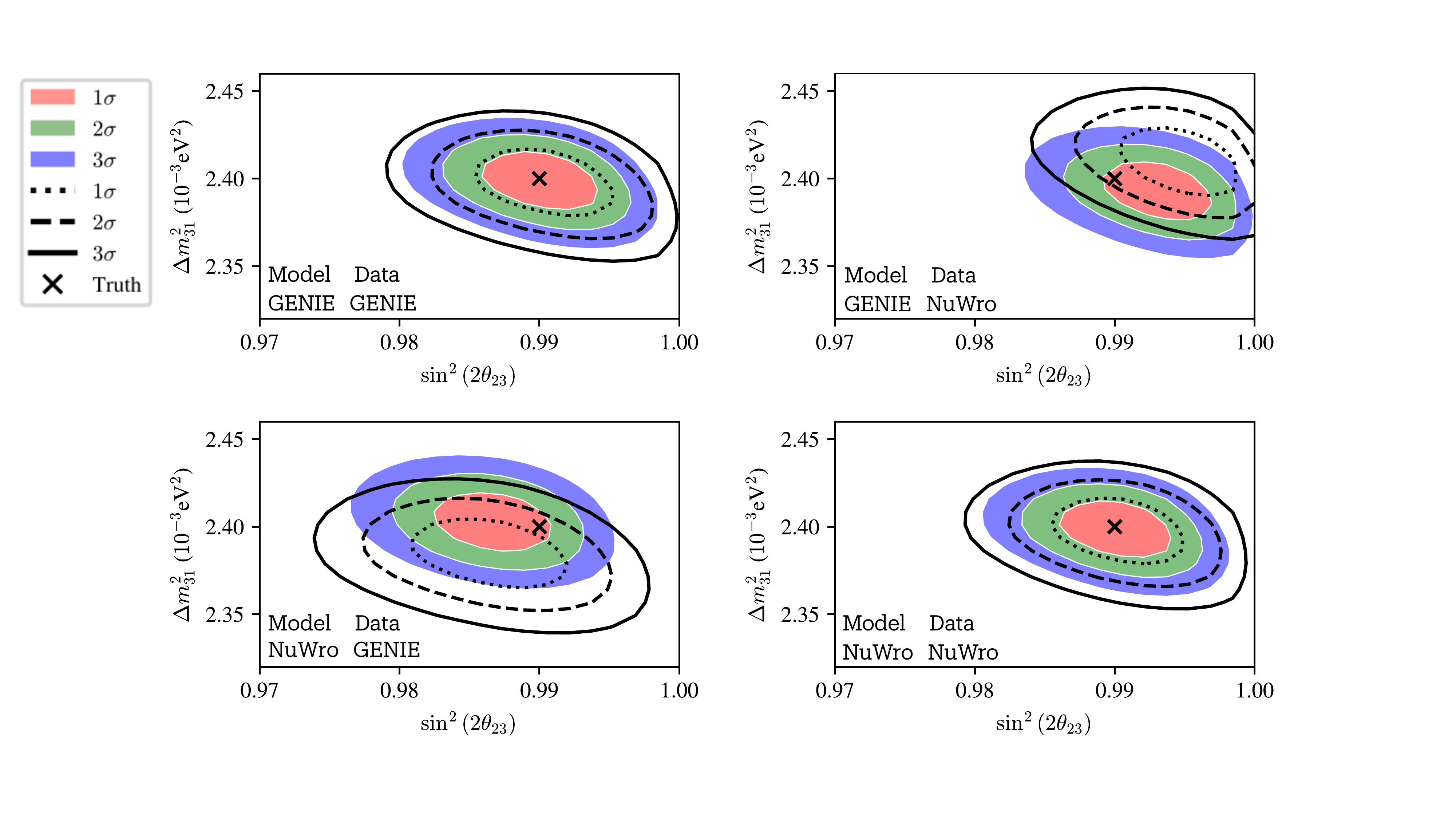}
\caption{
Approximate DUNE sensitivities to atmospheric oscillation parameters 
with (colored) and without (lines) the classifier. 
{\bf Top left:} \GENIE\ is used for mock data, fit, and classifier training. 
{\bf Bottom left:} \nuwro\ is used for mock data and \GENIE\ for fit and classifier training.  {\bf Top right:} \GENIE\ is used for mock data and \nuwro\ for fit and classifier training.
{\bf Bottom right:} \nuwro\ is used for mock data, fit, and classifier training. }
\label{fig:oscillation_sens_mismod_all_Ecal_only}
\end{center}
\end{figure*}

Figure~\ref{fig:cross_multi} examines generalization when each set of OvR classifiers learned on data from one generator is applied to data from the other.
Solid lines correspond to a classifier trained on \GENIE data and tested on \nuwro events, while dashed lines show the converse scenario. 
As before, classification of \nuwro data is typically better than that of \GENIE data. 
Nevertheless, we conclude that the classifiers generalize well. We note that, as in Fig.~\ref{fig:bin_multi}, there is a spike in contamination at near-zero efficiency in Fig.~\ref{fig:cross_multi}. 
However, it is due to a different reason. 
When the classifier is trained and tested on different generators, a systematic mismatch between the learned decision boundary and the test distribution leads to a non-zero false positive rate at all thresholds. As the true positive rate vanishes, this irreducible false positive contribution causes the contamination to spike, rather than the finite-statistics fluctuations responsible for the analogous feature in Fig.~\ref{fig:bin_multi}. Again this does not affect the following results.

Just as in the binary classification exercise of Sec.~\ref{sec:binary-class}, we observe that
the classifier models seem to perform according to the dataset they are tested on, rather than the dataset from which the classifier was learned.
Comparing Fig.~\ref{fig:cross_multi} with Fig.~\ref{fig:bin_multi}, we see that the rough shape of each curve is determined by the dataset being classified, rather than the model.
Deviations in performance between different classifier models on the same test dataset are more significant than in the QE vs.\ MEC case, especially near extremal choices of the threshold $\theta=0$ or $\theta=1$; however, shifts are relatively small for the intermediate values of $\theta$ which provide good performance.
As in the binary case, this suggests that the classifier is learning generic features of the kinematics of each interaction type, rather than unphysical features of the generator models.
This capacity for generalization is a necessary precondition for applying these methods to real experimental data where true interaction type labels are not available.
In the next section, we simulate such an application to an oscillation analysis at DUNE.


\section{Oscillation Analysis}
\label{sec:oscillation}

We turn now to examining how beneficial our classifier can be for neutrino oscillation parameter estimation.
As discussed in Sec.~\ref{sec:interaction}, different interaction types have different characteristic mechanisms for energy loss.
These lead to different neutrino energy reconstructions---i.e., different maps from the energy of the neutrino to the distribution of energies observed when that neutrino interacts with a nucleus.
Accurate constraints of neutrino oscillation parameters require this mapping to be well-characterized.
However, these effects depend on nuclear physics, and instead must be modeled.
Because truth labels are not available on real-world experimental data, it is difficult to separately constrain the systematics for each interaction type.
We demonstrate that using a learned classifier to (approximately) separate (mock) experimental data on interaction types provides useful additional constraints on these systematics, sufficient to enable a more accurate oscillation analysis.

To investigate, we conduct a toy analysis of DUNE's sensitivity to two key oscillation parameters, $\sin^2(2\theta_{23})$ and $\Delta m^2_{31}$.
For the mock data used as a proxy for experimental measurements, we assume $\sin^2\theta_{23}=0.45$ and $\Delta m^2_{31}=2.4\times10^{-3}$~eV$^2$.
We simulate an exposure of 480 kton-MW-years in neutrino mode (corresponding to, e.g.,~10 years of DUNE running with a 40 kton detector fiducial mass and a beam power of 1.2~MW).
The analysis considers only $\nu_\mu$ disappearance data and ignores backgrounds, as they are negligible for this channel.
We employ an analysis based on bins in calorimetric energy, see Eq.~\eqref{eq:calo}, from 0.1 to 10 GeV, using 200 MeV bins below 5 GeV and 500 MeV bins for higher energies.

We apply the learned classifier to each event to assign class labels $A$. 
In more detail, each event is assigned to the class for which the corresponding one-vs-rest binary classifier returns the highest signal probability, partitioning the event sample into four disjoint subsets that are independently binned in calorimetric energy.
The result is a set of event counts $N_j^A$ corresponding to the number of events in the calorimetric energy bin $j$ labeled as belonging to class $A$.
To estimate the impact of the classifier on oscillation measurements, we consider three types of uncertainties: overall normalization, spectral tilt, and energy scale. 
We implement these uncertainties as pull parameters ($x^A$, $y^A$, and $\Delta E^A$) in a $\chi^2$ test statistic.
The normalization and tilt are applied as
\begin{align}
    &\overline{N}_j^A = \left[1+x^A+y^A\left(1-\frac{2j}{n_{\rm bins}}\right)\right]N_j^A.
\end{align}
The energy scale uncertainties bias the calorimetric energies in each class independently by a constant given by the pull parameter $\Delta E^A$, redistributing events in reconstructed energy bins.
They are implemented as
\begin{equation}
    \mathcal{N}^A_j = \left( 1 - \left| \frac{\Delta E^A}{\Delta E_j} \right| \right) \overline{N}^A_j + 
    \begin{cases}
        \left| \frac{\Delta E^A}{\Delta E_{j+1}} \right|
        \overline{N}^A_{j+1}, & \Delta E^A < 0\\
        \left| \frac{\Delta E^A}{\Delta E_{j-1}} \right|
        \overline{N}^A_{j-1}, & \Delta E^A > 0
    \end{cases}
\end{equation}
where $\Delta E_j$ is the width of bin $j$.
In the fitting, we restrict $|\Delta E^A| \leq 200~\mathrm{MeV}$, the minimum bin width employed, thereby only allowing mixing between nearest-neighbor bins. For the left- or right-most bins, events are simply lost when shifted outside the range. 
In total, we have 12 uncertainties encoded as pull parameters.

Oscillated events from the generator chosen for the mock-data are binned to form the data, $D^A_j$, after first applying the classifier to assign class labels $A$.
The $\chi^2$ test statistic for the classifier case then reads
\begin{equation}
    \chi^2_{\rm class} = \sum_A^{\rm class} \sum_j^{\rm bins} \frac{(D_j^A - \mathcal{N}_j^A)^2}{\mathcal{N}_j^A} + {\rm priors},
\end{equation}
where $D_j^A$ is the mock data in calorimetric energy bin $j$ and class $A$, $\mathcal{N}_j^A$ is the modeled number of events after uncertainties, and ``priors'' are simply Gaussian priors of the form $(x^A/\sigma_x^A)^2$.
We assume 5\% priors for all normalization and tilt uncertainties, plus 100 MeV priors on the energy scale uncertainties, with no class dependence.

For the purpose of comparison, we use the same parametrization of uncertainties as in the ``no classifier'' fit,  but applied in combination independent of interaction label as
\begin{equation}
    \chi^2_\text{no class} = \sum_j^{\rm bins} \frac{(D_j - \mathcal{N}_j)^2}{\mathcal{N}_j} + {\rm priors},
\end{equation}
where $D_j$ is the mock data in calorimetric energy bin $j$ without any class labels, and $\mathcal{N}_j\equiv \sum_A^{\rm int} \mathcal{N}_j^A$ where the class labels are defined from the true interaction labels (QE, MEC, RES and DIS) taken from the generator used for the fit data, rather than from a classifier model. 
Note that labels are \emph{not} required for the experimental (mock) data, only the fit data.

Before proceeding to results, we caution that the sensitivities presented here are derived from a mock analysis designed to isolate the effect of event classification. 
Consequently, while our ``no classifier'' baseline is qualitatively similar to the DUNE nominal sensitivity, a direct quantitative comparison is not appropriate due to differences in the treatment of systematic uncertainties and analysis complexity. 
Therefore, the primary focus should be on the relative improvement gained by implementing our classifier, which demonstrates the method's potential to enhance future precision measurements. 

Figure~\ref{fig:oscillation_sens_mismod_all_Ecal_only} summarizes the results for the four different possible training and testing configurations.
In the case of no mismodeling, i.e.~when the same generator is used consistently throughout the analysis, the result is
a $\sim 10$-$20\%$ 
reduction in uncertainties without introducing any additional bias.
Given perfect modeling of neutrino-nucleus interactions, this approach may provide a modest but real improvement in the precision that can be obtained from finite experimental data.
However, perfect nuclear modeling will not be achieved in reality.

In a more realistic setting, both classifier and fit rely on the neutrino event generator which may not agree with data.
The crossed-generator results in Fig.~\ref{fig:oscillation_sens_mismod_all_Ecal_only} address this case.
In the baseline results without employing the classifier, we see that by using different generators for mock data and fitting leads to a clear bias on the estimate of the oscillation parameters, with the true oscillation parameters $\sim 1-2\sigma$ away from the best-fit (i.e., the center of hollow confidence intervals).
When using GENIE as mock data and NuWro for the fit the allowed region prefers lower $\Delta m^2_{31}$.
This is because the determination of the mass splitting is dominated by the energy corresponding to the minimum of oscillation. 
If GENIE is mock data and NuWro is used for the fit, since GENIE has more missing energy (see Fig.~\ref{fig:Emiss}), fitting with NuWro  leads to an apparent higher energy for the oscillation minimum. 
This translates to a lower value of the mass splitting.
The opposite happens when NuWro is taken for mock data and the fit is performed with GENIE.

The presence of a bias when using different generators of mock data and fit is not surprising; neutrino-nucleus interaction mismodeling is known to introduce such biases, which can be much more severe~\cite{Coyle:2025xjk,Coyle:2022bwa}.
Regardless, the tuning and the classification procedures can be carried out independently, e.g. as in the NOvA quartile analysis~\cite{NOvA:2018gge}.
We leave the exploration of the interplay between classification and tuning to future work.

The improvements from event-type classification in the mismodeling case is clearly apparent.
Significant improvements in sensitivity are only apparent in one example.
However, the reduction in bias is substantial in both tests.
Only the classifier-enabled results include the true parameters within the $1\sigma$ region, while the standard fit does so only at the $2\sigma$ level. 
This demonstrates that the approach is not just robust against mismodeling, but that it can make oscillation analyses more robust against mismodeling's biasing effects.

We note that several variations of the same method have achieved closely comparable performance:
\begin{itemize}[leftmargin=*]
\item Passing different input information, e.g.~adding the number of muons, protons, pions, and baryons per event, or e.g.~a simplified input excluding the per-particle calorimetric energy contribution, yields similar results in both classifier performance and oscillation analysis results.
We chose the setup for the main text as they were marginally better than the other options.
\item A simplified classification task: classifying events into soft interactions of QE and MEC, and hard interactions of RES and DIS. This is motivated by the large overlap in phase space between QE and MEC, and between RES and DIS. In addition, when NOvA first developed their quartile techniques, they evaluated schemes with 2, 4, 6, and 8 classes and found that the most significant improvement came from using two classes~\cite{Vinton:2018aqq}. Our observation is consistent with their conclusion.
\item The use of random forests~\cite{Breiman:2001hzm, biau2015randomforestguidedtour} rather than the multilayer perceptrons presented here. Random forests provide better performance for smaller datasets, but with a larger training dataset ($\sim 10^6$ events) we obtained similar results for all metrics (with NNs marginally better).
\end{itemize}


\section{Conclusions}
\label{sec:conclusions}

In this work, we have explored the use of supervised machine learning to classify neutrino events by their underlying interaction types using only final-state kinematics. This approach is motivated by the fact that different interaction mechanisms produce systematically different amounts of missing energy and therefore exhibit distinct calorimetric energy reconstruction performance. Standard analyses do not exploit this structure, motivating the use of classification to isolate event populations with more uniform reconstruction behavior.

Using simulated datasets from the \GENIE and \nuwro generators, we showed that the resulting classifiers achieve high efficiency and low contamination across interaction channels. Crucially, the classifiers generalize well across generators, indicating that they capture genuine kinematic signatures of the underlying microphysics rather than generator-specific artifacts. This suggests that classifier-based event labeling is robust against modeling differences and therefore suitable for application to real experimental data.

We integrated the classifier into a simplified DUNE $\nu_\mu$ disappearance analysis to quantify its impact on oscillation-parameter determination. We find that separating event classes yields tighter constraints on $\Delta m^2_{31}$ and $\sin^2(2\theta_{23})$. More importantly, event classification allowed more robust inference of oscillation parameters in the presence of nuclear physics mismodeling.
When tested in a cross-generator setup emulating an application to real experimental data,
classification lead to smaller confidence regions and less biases, demonstrating its potential to reduce reconstruction-driven systematic uncertainties.

Beyond the specific application considered here, the ability to separate events by their intrinsic interaction dynamics opens possibilities for improved cross-section tuning, model validation, and FD/ND extrapolation in long-baseline experiments. The method is also naturally compatible with analyses relying on kinematic energy reconstruction, such as those in T2K and Hyper-Kamiokande, where identifying the underlying interaction is an integral part of the reconstruction strategy.

Overall, our results indicate that incorporating interaction-type classification into the analysis pipeline offers a practical and robust path toward reducing systematic uncertainties and enhancing the precision of oscillation measurements in the high-statistics era of neutrino physics.


\section{Acknowledgments}
This document was prepared using the resources of the Fermi National Accelerator Laboratory (Fermilab), a U.S. Department of Energy, Office of Science, Office of High Energy Physics HEP User Facility. Fermilab is managed by Fermi Forward Discovery Group, LLC, acting under Contract No. 89243024CSC000002.
The work of K.T. is supported by DOE Grant KA2401045 (FNAL 23-32). The work of K.T. was performed in part at the Aspen Center for Physics, which is supported by a grant from the Alfred P. Sloan Foundation (G-2024-22395).
The work of SARE was supported in part
by SNF Ambizione grant PZ00P2\_193322, \textit{New frontiers
from sub-eV to super-TeV}.
This research was supported in part by grant NSF PHY-2309135 to the Kavli Institute for Theoretical Physics (KITP).

\bibliographystyle{apsrev4-1}
\bibliography{refs}

@article{DUNE:2020ypp,
    author = "Abi, Babak and others",
    collaboration = "DUNE",
    title = "{Deep Underground Neutrino Experiment (DUNE), Far Detector Technical Design Report, Volume II: DUNE Physics}",
    eprint = "2002.03005",
    archivePrefix = "arXiv",
    primaryClass = "hep-ex",
    reportNumber = "FERMILAB-PUB-20-025-ND, FERMILAB-DESIGN-2020-02",
    month = "2",
    year = "2020"
}

@article{T2K:2023mcm,
    author = "Abe, K. and others",
    collaboration = "T2K",
    title = "{Updated T2K measurements of muon neutrino and antineutrino disappearance using 3.6{\texttimes}1021 protons on target}",
    eprint = "2305.09916",
    archivePrefix = "arXiv",
    primaryClass = "hep-ex",
    doi = "10.1103/PhysRevD.108.072011",
    journal = "Phys. Rev. D",
    volume = "108",
    number = "7",
    pages = "072011",
    year = "2023"
}

@article{Hyper-Kamiokande:2018ofw,
    author = "Abe, K. and others",
    collaboration = "Hyper-Kamiokande",
    title = "{Hyper-Kamiokande Design Report}",
    eprint = "1805.04163",
    archivePrefix = "arXiv",
    primaryClass = "physics.ins-det",
    month = "5",
    year = "2018"
}

@article{Friedland:2018vry,
    author = "Friedland, Alexander and Li, Shirley Weishi",
    title = "{Understanding the energy resolution of liquid argon neutrino detectors}",
    eprint = "1811.06159",
    archivePrefix = "arXiv",
    primaryClass = "hep-ph",
    reportNumber = "SLAC-PUB-17352",
    doi = "10.1103/PhysRevD.99.036009",
    journal = "Phys. Rev. D",
    volume = "99",
    number = "3",
    pages = "036009",
    year = "2019"
}

@article{CLAS:2021neh,
    author = "Khachatryan, M. and others",
    collaboration = "CLAS, e4v",
    title = "{Electron-beam energy reconstruction for neutrino oscillation measurements}",
    doi = "10.1038/s41586-021-04046-5",
    journal = "Nature",
    volume = "599",
    number = "7886",
    pages = "565--570",
    year = "2021"
}

@article{Ankowski:2015kya,
    author = "Ankowski, Artur M. and Coloma, Pilar and Huber, Patrick and Mariani, Camillo and Vagnoni, Erica",
    title = "{Missing energy and the measurement of the CP-violating phase in neutrino oscillations}",
    eprint = "1507.08561",
    archivePrefix = "arXiv",
    primaryClass = "hep-ph",
    reportNumber = "FERMILAB-PUB-15-320-T",
    doi = "10.1103/PhysRevD.92.091301",
    journal = "Phys. Rev. D",
    volume = "92",
    number = "9",
    pages = "091301",
    year = "2015"
}

@article{NOvA:2021nfi,
    author = "Acero, M. A. and others",
    collaboration = "NOvA",
    title = "{Improved measurement of neutrino oscillation parameters by the NOvA experiment}",
    eprint = "2108.08219",
    archivePrefix = "arXiv",
    primaryClass = "hep-ex",
    reportNumber = "FERMILAB-PUB-21-373-ND",
    doi = "10.1103/PhysRevD.106.032004",
    journal = "Phys. Rev. D",
    volume = "106",
    number = "3",
    pages = "032004",
    year = "2022"
}

@article{Andreopoulos:2009rq,
    author = "Andreopoulos, C. and others",
    title = "{The GENIE Neutrino Monte Carlo Generator}",
    eprint = "0905.2517",
    archivePrefix = "arXiv",
    primaryClass = "hep-ph",
    reportNumber = "FERMILAB-PUB-09-418-CD",
    doi = "10.1016/j.nima.2009.12.009",
    journal = "Nucl. Instrum. Meth. A",
    volume = "614",
    pages = "87--104",
    year = "2010"
}

@article{Andreopoulos:2015wxa,
    author = "Andreopoulos, Costas and Barry, Christopher and Dytman, Steve and 
              Gallagher, Hugh and Golan, Tomasz and Hatcher, Robert and 
              Perdue, Gabriel and Yarba, Julia",
    title = "{The GENIE Neutrino Monte Carlo Generator: Physics and User Manual}",
    eprint = "1510.05494",
    archivePrefix = "arXiv",
    primaryClass = "hep-ph",
    reportNumber = "FERMILAB-FN-1004-CD",
    month = "10",
    year = "2015"
}

@article{Hayato:2009zz,
    author = "Hayato, Yoshinari",
    editor = "Ankowski, Arthur and Sobczyk, Jan",
    title = "{A neutrino interaction simulation program library NEUT}",
    journal = "Acta Phys. Polon. B",
    volume = "40",
    pages = "2477--2489",
    year = "2009"
}

@article{Buss:2011mx,
    author = "Buss, O. and Gaitanos, T. and Gallmeister, K. and van Hees, H. and Kaskulov, M. and Lalakulich, O. and Larionov, A. B. and Leitner, T. and Weil, J. and Mosel, U.",
    title = "{Transport-theoretical Description of Nuclear Reactions}",
    eprint = "1106.1344",
    archivePrefix = "arXiv",
    primaryClass = "hep-ph",
    doi = "10.1016/j.physrep.2011.12.001",
    journal = "Phys. Rept.",
    volume = "512",
    pages = "1--124",
    year = "2012"
}

@article{Golan:2012rfa,
    author = "Golan, T. and Sobczyk, J. T. and Zmuda, J.",
    editor = "Tzanakos, George S.",
    title = "{NuWro: the Wroclaw Monte Carlo Generator of Neutrino Interactions}",
    doi = "10.1016/j.nuclphysbps.2012.09.136",
    journal = "Nucl. Phys. B Proc. Suppl.",
    volume = "229-232",
    pages = "499--499",
    year = "2012"
}

@article{Gallmeister:2016dnq,
    author = "Gallmeister, K. and Mosel, U. and Weil, J.",
    title = "{Neutrino-Induced Reactions on Nuclei}",
    eprint = "1605.09391",
    archivePrefix = "arXiv",
    primaryClass = "nucl-th",
    doi = "10.1103/PhysRevC.94.035502",
    journal = "Phys. Rev. C",
    volume = "94",
    number = "3",
    pages = "035502",
    year = "2016"
}

@article{Isaacson:2022cwh,
    author = "Isaacson, Joshua and Jay, William I. and Lovato, Alessandro and Machado, Pedro A. N. and Rocco, Noemi",
    title = "{Introducing a novel event generator for electron-nucleus and neutrino-nucleus scattering}",
    eprint = "2205.06378",
    archivePrefix = "arXiv",
    primaryClass = "hep-ph",
    reportNumber = "FERMILAB-PUB-22-411-T, MIT-CTP/5428",
    doi = "10.1103/PhysRevD.107.033007",
    journal = "Phys. Rev. D",
    volume = "107",
    number = "3",
    pages = "033007",
    year = "2023"
}

@article{NuSTEC:2017hzk,
    author = "Alvarez-Ruso, L. and others",
    collaboration = "NuSTEC",
    title = "{NuSTEC White Paper: Status and challenges of neutrino\textendash{}nucleus scattering}",
    eprint = "1706.03621",
    archivePrefix = "arXiv",
    primaryClass = "hep-ph",
    reportNumber = "FERMILAB-PUB-17-195-ND-T, INT-PUB-17-020",
    doi = "10.1016/j.ppnp.2018.01.006",
    journal = "Prog. Part. Nucl. Phys.",
    volume = "100",
    pages = "1--68",
    year = "2018"
}

@article{Kronfeld:2019nfb,
    author = "Kronfeld, Andreas S. and Richards, David G. and Detmold, William and Gupta, Rajan and Lin, Huey-Wen and Liu, Keh-Fei and Meyer, Aaron S. and Sufian, Raza and Syritsyn, Sergey",
    collaboration = "USQCD",
    title = "{Lattice QCD and Neutrino-Nucleus Scattering}",
    eprint = "1904.09931",
    archivePrefix = "arXiv",
    primaryClass = "hep-lat",
    reportNumber = "FERMILAB-PUB-19-172-T",
    doi = "10.1140/epja/i2019-12916-x",
    journal = "Eur. Phys. J. A",
    volume = "55",
    number = "11",
    pages = "196",
    year = "2019"
}

@article{Ruso:2022qes,
    author = "Ruso, L. Alvarez and others",
    title = "{Theoretical tools for neutrino scattering: interplay between lattice QCD, EFTs, nuclear physics, phenomenology, and neutrino event generators}",
    eprint = "2203.09030",
    archivePrefix = "arXiv",
    primaryClass = "hep-ph",
    reportNumber = "DESY-22-05, FERMILAB-FN-1161-T, MITP-22-027, DESY-22-052",
    month = "3",
    year = "2022"
}

@article{Bloom:1970xb,
    author = "Bloom, Elliott D. and Gilman, Frederick J.",
    title = "{Scaling, Duality, and the Behavior of Resonances in Inelastic electron-Proton Scattering}",
    reportNumber = "SLAC-PUB-0779",
    doi = "10.1103/PhysRevLett.25.1140",
    journal = "Phys. Rev. Lett.",
    volume = "25",
    pages = "1140",
    year = "1970"
}

@article{Bodek:2002vp,
    author = "Bodek, A. and Yang, U. K.",
    editor = "Morfin, J. G. and Sakuda, M. and Suzuki, Y.",
    title = "{Modeling deep inelastic cross-sections in the few GeV region}",
    eprint = "hep-ex/0203009",
    archivePrefix = "arXiv",
    doi = "10.1016/S0920-5632(02)01755-3",
    journal = "Nucl. Phys. B Proc. Suppl.",
    volume = "112",
    pages = "70--76",
    year = "2002"
}

@article{Ankowski:2005wi,
    author = "Ankowski, Artur M. and Sobczyk, Jan T.",
    title = "{Argon spectral function and neutrino interactions}",
    eprint = "nucl-th/0512004",
    archivePrefix = "arXiv",
    doi = "10.1103/PhysRevC.74.054316",
    journal = "Phys. Rev. C",
    volume = "74",
    pages = "054316",
    year = "2006"
}

@article{Drechsel:2007if,
    author = "Drechsel, D. and Kamalov, S. S. and Tiator, L.",
    title = "{Unitary Isobar Model - MAID2007}",
    eprint = "0710.0306",
    archivePrefix = "arXiv",
    primaryClass = "nucl-th",
    doi = "10.1140/epja/i2007-10490-6",
    journal = "Eur. Phys. J. A",
    volume = "34",
    pages = "69--97",
    year = "2007"
}

@article{Bodek:2010km,
    author = "Bodek, Arie and Yang, Un-ki",
    title = "{Axial and Vector Structure Functions for Electron- and Neutrino- Nucleon Scattering Cross Sections at all $Q^2$ using Effective Leading order Parton Distribution Functions}",
    eprint = "1011.6592",
    archivePrefix = "arXiv",
    primaryClass = "hep-ph",
    month = "11",
    year = "2010"
}

@article{Kamano:2013iva,
    author = "Kamano, H. and Nakamura, S. X. and Lee, T. -S. H. and Sato, T.",
    title = "{Nucleon resonances within a dynamical coupled-channels model of $\pi N$ and $\gamma N$ reactions}",
    eprint = "1305.4351",
    archivePrefix = "arXiv",
    primaryClass = "nucl-th",
    doi = "10.1103/PhysRevC.88.035209",
    journal = "Phys. Rev. C",
    volume = "88",
    number = "3",
    pages = "035209",
    year = "2013"
}

@article{Nakamura:2015rta,
    author = "Nakamura, S. X. and Kamano, H. and Sato, T.",
    title = "{Dynamical coupled-channels model for neutrino-induced meson productions in resonance region}",
    eprint = "1506.03403",
    archivePrefix = "arXiv",
    primaryClass = "hep-ph",
    doi = "10.1103/PhysRevD.92.074024",
    journal = "Phys. Rev. D",
    volume = "92",
    number = "7",
    pages = "074024",
    year = "2015"
}

@article{Kabirnezhad:2017jmf,
    author = "Kabirnezhad, Monireh",
    title = "{Single pion production in neutrino-nucleon Interactions}",
    eprint = "1711.02403",
    archivePrefix = "arXiv",
    primaryClass = "hep-ph",
    doi = "10.1103/PhysRevD.97.013002",
    journal = "Phys. Rev. D",
    volume = "97",
    number = "1",
    pages = "013002",
    year = "2018"
}

@article{Xie:2023suk,
    author = "Xie, Keping and Gao, Jun and Hobbs, T. J. and Stump, Daniel R. and Yuan, C. -P.",
    title = "{High-energy neutrino deeply inelastic scattering cross sections from 100 GeV to 1000 EeV}",
    eprint = "2303.13607",
    archivePrefix = "arXiv",
    primaryClass = "hep-ph",
    reportNumber = "ANL-181430, MSUHEP-23-003, PITT-PACC-2302",
    month = "3",
    year = "2023"
}

@article{Coyle:2025xjk,
    author = "Coyle, Nina M. and Li, Shirley Weishi and Machado, Pedro A. N.",
    title = "{Neutrino-nucleus cross section impacts on neutrino oscillation measurements}",
    eprint = "2502.19467",
    archivePrefix = "arXiv",
    primaryClass = "hep-ph",
    reportNumber = "FERMILAB-PUB-25-0001-T, UCI-HEP-TR-2025-01",
    doi = "10.1103/PhysRevD.111.093010",
    journal = "Phys. Rev. D",
    volume = "111",
    number = "9",
    pages = "093010",
    year = "2025"
}

@misc{DUNE_fluxes,
    author = "Fields, Laura",
    title = "{DUNE Fluxes}",
    note = {\url{https://glaucus.crc.nd.edu/DUNEFluxes/}},
}

@article{Coyle:2022bwa,
    author = "Coyle, Nina M. and Li, Shirley Weishi and Machado, Pedro A. N.",
    title = "{The impact of neutrino-nucleus interaction modeling on new physics searches}",
    eprint = "2210.03753",
    archivePrefix = "arXiv",
    primaryClass = "hep-ph",
    reportNumber = "FERMILAB-PUB-22-726-T",
    doi = "10.1007/JHEP12(2022)166",
    journal = "JHEP",
    volume = "12",
    pages = "166",
    year = "2022"
}

@article{Breiman:2001hzm,
    author = "Breiman, Leo",
    title = "{Random Forests}",
    doi = "10.1023/A:1010933404324",
    journal = "Machine Learning",
    volume = "45",
    number = "1",
    pages = "5--32",
    year = "2001"
}

@misc{biau2015randomforestguidedtour,
      title={A Random Forest Guided Tour}, 
      author={Gérard Biau and Erwan Scornet},
      year={2015},
      eprint={1511.05741},
      archivePrefix={arXiv},
      primaryClass={math.ST},
      url={https://arxiv.org/abs/1511.05741}, 
}

@article{Hornik:1989yye,
    author = "Hornik, Kurt and Stinchcombe, Maxwell and White, Halbert",
    title = "{Multilayer feedforward networks are universal approximators}",
    doi = "10.1016/0893-6080(89)90020-8",
    journal = "Neural Networks",
    volume = "2",
    number = "5",
    pages = "359--366",
    year = "1989"
}

@article{Hinton:2012zps,
    author = "Hinton, Geoffrey E. and Srivastava, Nitish and Krizhevsky, Alex and Sutskever, Ilya and Salakhutdinov, Ruslan R.",
    title = "{Improving neural networks by preventing co-adaptation of feature detectors}",
    eprint = "1207.0580",
    archivePrefix = "arXiv",
    primaryClass = "cs.NE",
    month = "7",
    year = "2012"
}

@article{scikit-learn,
  title={Scikit-learn: Machine Learning in {P}ython},
  author={Pedregosa, F. and Varoquaux, G. and Gramfort, A. and Michel, V.
          and Thirion, B. and Grisel, O. and Blondel, M. and Prettenhofer, P.
          and Weiss, R. and Dubourg, V. and Vanderplas, J. and Passos, A. and
          Cournapeau, D. and Brucher, M. and Perrot, M. and Duchesnay, E.},
  journal={Journal of Machine Learning Research},
  volume={12},
  pages={2825--2830},
  year={2011}
}

@article{NOvA:2018gge,
    author = "Acero, M. A. and others",
    collaboration = "NOvA",
    title = "{New constraints on oscillation parameters from $\nu_e$ appearance and $\nu_\mu$ disappearance in the NOvA experiment}",
    eprint = "1806.00096",
    archivePrefix = "arXiv",
    primaryClass = "hep-ex",
    reportNumber = "FERMILAB-PUB-18-223-ND",
    doi = "10.1103/PhysRevD.98.032012",
    journal = "Phys. Rev. D",
    volume = "98",
    pages = "032012",
    year = "2018"
}

@phdthesis{Vinton:2018aqq,
    author = "Vinton, Luke",
    title = "{Measurement of muon neutrino disappearance with a NOvA experiment}",
    reportNumber = "FERMILAB-THESIS-2018-05",
    doi = "10.2172/1423216",
    school = "Sussex U., Sussex U.",
    year = "2018"
}

@article{MicroBooNE:2024hun,
    author = "Abratenko, P. and others",
    collaboration = "MicroBooNE",
    title = "{Demonstration of neutron identification in neutrino interactions in the MicroBooNE liquid argon time projection chamber}",
    eprint = "2406.10583",
    archivePrefix = "arXiv",
    primaryClass = "hep-ex",
    reportNumber = "FERMILAB-PUB-24-0301",
    doi = "10.1140/epjc/s10052-024-13423-z",
    journal = "Eur. Phys. J. C",
    volume = "84",
    number = "10",
    pages = "1052",
    year = "2024"
}

@article{ARTIE:2022wqs,
    author = "Andringa, S. and others",
    collaboration = "ARTIE",
    title = "{Measurement of the total neutron cross section~on argon in the 20 to 70 keV energy range}",
    eprint = "2212.05448",
    archivePrefix = "arXiv",
    primaryClass = "nucl-ex",
    doi = "10.1103/PhysRevC.108.L011601",
    journal = "Phys. Rev. C",
    volume = "108",
    number = "1",
    pages = "L011601",
    year = "2023"
}

@article{CAPTAIN:2019fxo,
    author = "Bhandari, B. and others",
    collaboration = "CAPTAIN",
    title = "{First Measurement of the Total Neutron Cross Section on Argon Between 100 and 800 MeV}",
    eprint = "1903.05276",
    archivePrefix = "arXiv",
    primaryClass = "hep-ex",
    reportNumber = "LANL Report LA-UR-19-22200",
    doi = "10.1103/PhysRevLett.123.042502",
    journal = "Phys. Rev. Lett.",
    volume = "123",
    number = "4",
    pages = "042502",
    year = "2019"
}

@article{CAPTAIN:2022nzf,
    author = "Martynenko, S. and others",
    collaboration = "CAPTAIN",
    title = "{Measurement of the neutron cross section on argon between 95 and 720~MeV}",
    eprint = "2209.13488",
    archivePrefix = "arXiv",
    primaryClass = "nucl-ex",
    doi = "10.1103/PhysRevD.107.072009",
    journal = "Phys. Rev. D",
    volume = "107",
    number = "7",
    pages = "072009",
    year = "2023"
}

@article{ArgoNeuT:2018tvi,
    author = "Acciarri, R. and others",
    collaboration = "ArgoNeuT",
    title = "{Demonstration of MeV-Scale Physics in Liquid Argon Time Projection Chambers Using ArgoNeuT}",
    eprint = "1810.06502",
    archivePrefix = "arXiv",
    primaryClass = "hep-ex",
    reportNumber = "FERMILAB-PUB-18-559-ND",
    doi = "10.1103/PhysRevD.99.012002",
    journal = "Phys. Rev. D",
    volume = "99",
    number = "1",
    pages = "012002",
    year = "2019"
}

@article{Castiglioni:2020tsu,
    author = "Castiglioni, W. and Foreman, W. and Lepetic, I. and Littlejohn, B. R. and Malaker, M. and Mastbaum, A.",
    title = "{Benefits of MeV-scale reconstruction capabilities in large liquid argon time projection chambers}",
    eprint = "2006.14675",
    archivePrefix = "arXiv",
    primaryClass = "physics.ins-det",
    doi = "10.1103/PhysRevD.102.092010",
    journal = "Phys. Rev. D",
    volume = "102",
    number = "9",
    pages = "092010",
    year = "2020"
}

@inproceedings{Kingma:2014vow,
    author = "Kingma, Diederik P. and Ba, Jimmy",
    title = "{Adam: A Method for Stochastic Optimization}",
    eprint = "1412.6980",
    archivePrefix = "arXiv",
    primaryClass = "cs.LG",
    month = "12",
    year = "2014"
}

@article{DUNE:2015lol,
    author = "Acciarri, R. and others",
    collaboration = "DUNE",
    title = "{Long-Baseline Neutrino Facility (LBNF) and Deep Underground Neutrino Experiment (DUNE)}: {Conceptual Design Report, Volume 2: The Physics Program for DUNE at LBNF}",
    eprint = "1512.06148",
    archivePrefix = "arXiv",
    primaryClass = "physics.ins-det",
    reportNumber = "FERMILAB-DESIGN-2016-02",
    month = "12",
    year = "2015"
}

@inproceedings{Yu:2019yuj,
    author = "Yu, Shiqi",
    title = "{Electron Neutrino Energy Reconstruction in NOvA Using CNN Particle IDs}",
    booktitle = "{Meeting of the Division of Particles and Fields of the American Physical Society}",
    eprint = "1910.06953",
    archivePrefix = "arXiv",
    primaryClass = "physics.ins-det",
    month = "10",
    year = "2019"
}

@article{Liu:2020pzv,
    author = "Liu, Junze and Ott, Jordan and Collado, Julian and Jargowsky, Benjamin and Wu, Wenjie and Bian, Jianming and Baldi, Pierre",
    collaboration = "DUNE",
    title = "{Deep-Learning-Based Kinematic Reconstruction for DUNE}",
    eprint = "2012.06181",
    archivePrefix = "arXiv",
    primaryClass = "physics.ins-det",
    month = "12",
    year = "2020"
}

@article{Shmakov:2023jms,
    author = "Shmakov, Alexander and Yankelevich, Alejandro and Bian, Jianming and Baldi, Pierre",
    collaboration = "NOvA",
    title = "{Interpretable Joint Event-Particle Reconstruction for Neutrino Physics at NOvA with Sparse CNNs and Transformers}",
    eprint = "2303.06201",
    archivePrefix = "arXiv",
    primaryClass = "cs.LG",
    reportNumber = "FERMILAB-CONF-23-016-ND",
    month = "3",
    year = "2023"
}

@article{IceCube:2025jmv,
    author = "Abbasi, R. and others",
    collaboration = "IceCube",
    title = "{Fast Low Energy Reconstruction using Convolutional Neural Networks}",
    eprint = "2505.16777",
    archivePrefix = "arXiv",
    primaryClass = "astro-ph.HE",
    month = "5",
    year = "2025"
}

@article{Baldi:2018qhe,
    author = "Baldi, Pierre and Bian, Jianming and Hertel, Lars and Li, Lingge",
    title = "{Improved Energy Reconstruction in NOvA with Regression Convolutional Neural Networks}",
    eprint = "1811.04557",
    archivePrefix = "arXiv",
    primaryClass = "physics.ins-det",
    doi = "10.1103/PhysRevD.99.012011",
    journal = "Phys. Rev. D",
    volume = "99",
    number = "1",
    pages = "012011",
    year = "2019"
}

@article{MicroBooNE:2021pvo,
    author = "Abratenko, P. and others",
    collaboration = "MicroBooNE",
    title = "{Search for an anomalous excess of charged-current quasielastic {\ensuremath{\nu}}e interactions with the MicroBooNE experiment using Deep-Learning-based reconstruction}",
    eprint = "2110.14080",
    archivePrefix = "arXiv",
    primaryClass = "hep-ex",
    reportNumber = "FERMILAB-PUB-21-507-ND",
    doi = "10.1103/PhysRevD.105.112003",
    journal = "Phys. Rev. D",
    volume = "105",
    number = "11",
    pages = "112003",
    year = "2022"
}

@article{Friedland:2020cdp,
    author = "Friedland, Alexander and Li, Shirley Weishi",
    title = "{Simulating hadron test beams in liquid argon}",
    eprint = "2007.13336",
    archivePrefix = "arXiv",
    primaryClass = "hep-ph",
    reportNumber = "SLAC-PUB-17549",
    doi = "10.1103/PhysRevD.102.096005",
    journal = "Phys. Rev. D",
    volume = "102",
    number = "9",
    pages = "096005",
    year = "2020"
}

@inproceedings{Shifman:2000jv,
    author = "Shifman, Mikhail A.",
    title = "{Quark hadron duality}",
    booktitle = "{8th International Symposium on Heavy Flavor Physics}",
    eprint = "hep-ph/0009131",
    archivePrefix = "arXiv",
    reportNumber = "TPI-MINN-00-44, UMN-TH-1920-00",
    doi = "10.1142/9789812810458_0032",
    publisher = "World Scientific",
    address = "Singapore",
    volume = "3",
    pages = "1447--1494",
    month = "7",
    year = "2000"
}

@article{Melnitchouk:2005zr,
    author = "Melnitchouk, W. and Ent, R. and Keppel, C.",
    title = "{Quark-hadron duality in electron scattering}",
    eprint = "hep-ph/0501217",
    archivePrefix = "arXiv",
    reportNumber = "JLAB-THY-04-266",
    doi = "10.1016/j.physrep.2004.10.004",
    journal = "Phys. Rept.",
    volume = "406",
    pages = "127--301",
    year = "2005"
}

@article{Lalakulich:2006yn,
    author = "Lalakulich, O. and Melnitchouk, W. and Paschos, E. A.",
    title = "{Quark-hadron duality in neutrino scattering}",
    eprint = "hep-ph/0608058",
    archivePrefix = "arXiv",
    reportNumber = "DO-TH-05-09, JLAB-THY-06-515",
    doi = "10.1103/PhysRevC.75.015202",
    journal = "Phys. Rev. C",
    volume = "75",
    pages = "015202",
    year = "2007"
}

@article{Lalakulich:2008tu,
    author = "Lalakulich, Olga and Jachowicz, Natalie and Praet, Christophe and Ryckebusch, Jan",
    title = "{Quark-hadron duality in lepton scattering off nuclei}",
    eprint = "0808.0085",
    archivePrefix = "arXiv",
    primaryClass = "nucl-th",
    doi = "10.1103/PhysRevC.79.015206",
    journal = "Phys. Rev. C",
    volume = "79",
    pages = "015206",
    year = "2009"
}

@article{Bigi:2001ys,
    author = "Bigi, Ikaros I. Y. and Uraltsev, Nikolai",
    title = "{A Vademecum on quark hadron duality}",
    eprint = "hep-ph/0106346",
    archivePrefix = "arXiv",
    reportNumber = "UND-HEP-01-BIG-01, BICOCCA-FT-01-18, UND-HEP-01-BIG01",
    doi = "10.1142/S0217751X01005535",
    journal = "Int. J. Mod. Phys. A",
    volume = "16",
    pages = "5201--5248",
    year = "2001"
}

@article{Benhar:2005dj,
    author = "Benhar, Omar and Farina, Nicola and Nakamura, Hiroki and Sakuda, Makoto and Seki, Ryoichi",
    title = "{Electron- and neutrino-nucleus scattering in the impulse approximation regime}",
    eprint = "hep-ph/0506116",
    archivePrefix = "arXiv",
    doi = "10.1103/PhysRevD.72.053005",
    journal = "Phys. Rev. D",
    volume = "72",
    pages = "053005",
    year = "2005"
}

@article{Nieves:2011pp,
    author = "Nieves, J. and Ruiz Simo, I. and Vicente Vacas, M. J.",
    title = "{Inclusive Charged--Current Neutrino--Nucleus Reactions}",
    eprint = "1102.2777",
    archivePrefix = "arXiv",
    primaryClass = "hep-ph",
    doi = "10.1103/PhysRevC.83.045501",
    journal = "Phys. Rev. C",
    volume = "83",
    pages = "045501",
    year = "2011"
}

@article{LlewellynSmith:1971uhs,
    author = "Llewellyn Smith, C. H.",
    title = "{Neutrino Reactions at Accelerator Energies}",
    reportNumber = "SLAC-PUB-0958",
    doi = "10.1016/0370-1573(72)90010-5",
    journal = "Phys. Rept.",
    volume = "3",
    pages = "261--379",
    year = "1972"
}

@article{Rein:1980wg,
    author = "Rein, Dieter and Sehgal, Lalit M.",
    title = "{Neutrino Excitation of Baryon Resonances and Single Pion Production}",
    reportNumber = "PITHA-80-10",
    doi = "10.1016/0003-4916(81)90242-6",
    journal = "Annals Phys.",
    volume = "133",
    pages = "79--153",
    year = "1981"
}

@article{Martini:2009uj,
    author = "Martini, M. and Ericson, M. and Chanfray, G. and Marteau, J.",
    title = "{A Unified approach for nucleon knock-out, coherent and incoherent pion production in neutrino interactions with nuclei}",
    eprint = "0910.2622",
    archivePrefix = "arXiv",
    primaryClass = "nucl-th",
    reportNumber = "LYCEN-2009-10, CERN-PH-TH-2009-192",
    doi = "10.1103/PhysRevC.80.065501",
    journal = "Phys. Rev. C",
    volume = "80",
    pages = "065501",
    year = "2009"
}

@article{Rein:1982pf,
    author = "Rein, Dieter and Sehgal, Lalit M.",
    title = "{Coherent pi0 Production in Neutrino Reactions}",
    reportNumber = "PITHA-82-27",
    doi = "10.1016/0550-3213(83)90090-1",
    journal = "Nucl. Phys. B",
    volume = "223",
    pages = "29--44",
    year = "1983"
}

@article{MicroBooNE:2016pwy,
    author = "Acciarri, R. and others",
    collaboration = "MicroBooNE",
    title = "{Design and Construction of the MicroBooNE Detector}",
    eprint = "1612.05824",
    archivePrefix = "arXiv",
    primaryClass = "physics.ins-det",
    reportNumber = "FERMILAB-PUB-16-613-ND",
    doi = "10.1088/1748-0221/12/02/P02017",
    journal = "JINST",
    volume = "12",
    number = "02",
    pages = "P02017",
    year = "2017"
}

@article{MicroBooNE:2020akw,
    author = "Abratenko, P. and others",
    collaboration = "MicroBooNE",
    title = "{Measurement of differential cross sections for $\nu_{\mu}$ -Ar charged-current interactions with protons and no pions in the final state with the MicroBooNE detector}",
    eprint = "2010.02390",
    archivePrefix = "arXiv",
    primaryClass = "hep-ex",
    reportNumber = "FERMILAB-PUB-20-505-AD-ND-SCD-TD",
    doi = "10.1103/PhysRevD.102.112013",
    journal = "Phys. Rev. D",
    volume = "102",
    number = "11",
    pages = "112013",
    year = "2020"
}

\end{document}